\documentstyle[aps,prl,epsf]{revtex}
\begin{document}
\twocolumn[\hsize\textwidth\columnwidth
\hsize\csname@twocolumnfalse\endcsname
\draft

\title{Quantum phase interference (Berry phase) 
in single-molecule magnets of [Mn$_{12}$]$^{2-}$}

\author{W. Wernsdorfer$^1$, M. Soler$^2$, 
G. Christou$^2$, D.N. Hendrickson$^3$}
\address{
$^1$Lab. L. N\'eel, associ\'e \`a l'UJF, CNRS, BP 166,
38042 Grenoble Cedex 9, France\\
$^2$Dept. of Chemistry and Molecular Structure Center, Indiana Uni., 
Bloomington, 47405-4001, IN, USA\\
$^3$Dept. of Chemistry and Biochemistry, Uni. of California at San 
Diego, La Jolla, 92037, CA, USA
}
\date{Abstract Designation: {\bf BG-06}; version \today}
\maketitle
\begin{abstract}
Magnetization measurements of a molecular clusters 
(Mn$_{12}$)$^{2-}$ with a spin ground state of $S = 10$ show
resonance tunneling at avoided energy level crossings. 
The observed oscillations of the tunnel probability as a 
function of the magnetic field applied along the 
hard anisotropy axis are due to topological 
quantum phase interference of two tunnel paths of 
opposite windings. (Mn$_{12}$)$^{2-}$ is therefore the
second molecular clusters presenting 
quantum phase interference.

\end{abstract}
\bigskip
\vskip1pc]
\narrowtext

Studying the limits between classical and 
quantum physics has become a very attractive 
field of research. Single-molecule magnets (SMMs) are 
among the most promising candidates to observe 
these phenomena since they have a well defined 
structure with well characterized spin ground state 
and magnetic anisotropy~\cite{Aubin98,Caneschi99}. 
Quantum phase interference~\cite{Garg93} is among the most 
interesting quantum phenomena that can be studied
at the mesoscopic level in SMMs.
This effect was recently observed in the Fe$_8$ SMM~\cite{WW_Science99}.
It has led to new theoretical studies on quantum phase interference in spin systems 
~\cite{Garg99a,Garg99b,Garg99c,Garg00b,Garg00d,Barnes99,Villain00,Liang00,Yoo00,Yoo00b,Leuenberger00,Leuenberger01b,Lu00a,Lu00b,Lu00c,Zhang99,Jin00,Chudnovsky00a}.
We present here a second SMM, called [Mn$_{12}$]$^{2-}$, that clearly shows 
quantum phase interference effects.

\begin{figure}
\centerline{\epsfxsize=8 cm \epsfbox{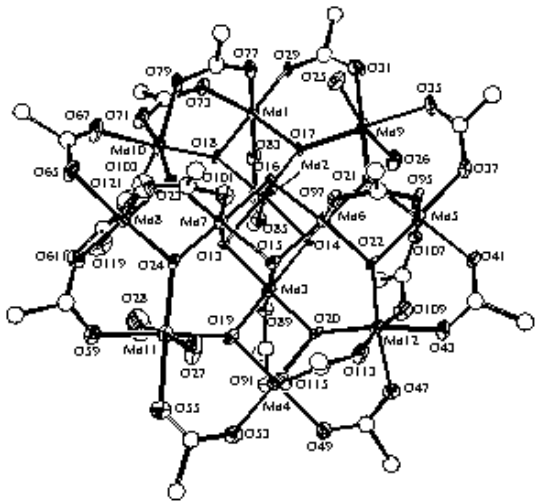}}
\caption{ORTEP representation of the complex anion 
[Mn$_{12}$O$_{12}$(O$_{2}$CCHCl$_{2}$)$_{16}$(H$_{2}$O)$_{4}$]$^{2-}$ 
showing 50 percent probability ellipsoids. The white open circles
represent C atoms.
For clarity, all H and Cl atoms have been omitted.}
\label{mol}
\end{figure}

The [Mn$_{12}$]$^{2-}$ single-molecule magnet was prepared from 
[Mn$_{12}$O$_{12}$(O$_{2}$Ac)$_{16}$(H$_{2}$O)$_{4}$] 
(Mn$_{12}$ acetate)~\cite{Caneschi99} by a ligand 
substitution procedure and converted to 
(PPh$_{4}$)$_{2}$[Mn$_{12}$O$_{12}$(O$_{2}$CCHCl$_{2}$)$_{16}$(H$_{2}$O)$_{4}$] 
by the two-electrons reduction with two 
equivalents of PPh$_4^+$I$^-$. 
The crystal structure (Fig. 1) shows that 
the two added electrons are on Mn$^{2+}$ ions 
Mn9 and Mn11 giving 
a 2 Mn$^{2+}$, 6 Mn$^{3+}$, 4 Mn$^{4+}$ description~\cite{Soler00}. 
The compound has a $S = 10$ spin ground state 
and negative (Ising type) magnetoanisotropy.

The simplest model describing the spin system of [Mn$_{12}$]$^{2-}$ 
has the following Hamiltonian
\begin{equation}
	H = -D S_z^2 + E \left(S_x^2 - S_y^2\right) 
	+ g \mu_{\rm B} \mu_0 \vec{S}\cdot\vec{H} 
\label{eq_H_biax}
\end{equation}
$S_x$, $S_y$, and $S_z$ are the three 
components of the spin operator, 
$D$  and $E$ are the anisotropy constants, 
and the last term describes the Zeeman 
energy associated with an applied field $\vec{H}$. 
This Hamiltonian defines hard, medium, 
and easy axes of magnetization in $x$, $y$, and $z$ directions, 
respectively. 
It has an energy level 
spectrum with $(2S+1) = 21$ values which, 
to a first approximation, can be labeled by the 
quantum numbers $m = -10, -9, ..., 10$
taking the $z$-axis as the quantization axis.The energy spectrum  
can be obtained by using 
standard diagonalisation techniques of the $[21 \times 21]$ 
matrix. At $\vec{H} = 0$, the levels $m = \pm 10$ 
have the lowest energy. 
When a field $H_z$ is applied, the levels with 
$m < 0$ increase in energy, while those 
with $m > 0$ decrease. Therefore, 
energy levels of positive and negative 
quantum numbers cross at certain values of $H_z$. 
given by $\mu_0 H_z \approx n D/g \mu_{\rm B}$, 
with $n = 0, 1, 2, 3, ...$. 

When the spin Hamiltonian contains transverse terms 
(for instance $E(S_x^2 - S_y^2)$), the level crossings  can be 
``avoided level crossings''. 
The spin $S$ is ``in resonance'' between two 
states when the local longitudinal field is close 
to an avoided level crossing. 
The energy gap, the so-called 
``tunnel spitting'' $\Delta$, can be tuned by 
a transverse field (a field applied perpendicular 
to the $z -$ direction) 
via the $S_xH_x$ and $S_yH_y$ Zeeman terms.
In the case of the transverse term $E(S_x^2 - S_y^2)$,
it was shown that $\Delta$ oscillates with
a period given by \cite{Garg93}
\begin{equation}
\Delta H = \frac {2 k_{\rm B}}{g \mu_{\rm B}} \sqrt{2 E (E + D)}
\label{eq_Garg}
\end{equation}
The oscillations are explained by constructive 
or destructive interference of quantum 
spin phases (Berry phases) of two tunnel paths \cite{Garg93}.

All measurements were performed using an array of 
micro-SQUIDs~\cite{WW_PRL99}. The high sensitivity of this 
magnetometer allows the study of single crystals of SMMs
with sizes of the order of 10 to 500 $\mu$m.
The field can be applied in any direction by separately 
driving three orthogonal coils.

\begin{figure}
\centerline{\epsfxsize=8 cm \epsfbox{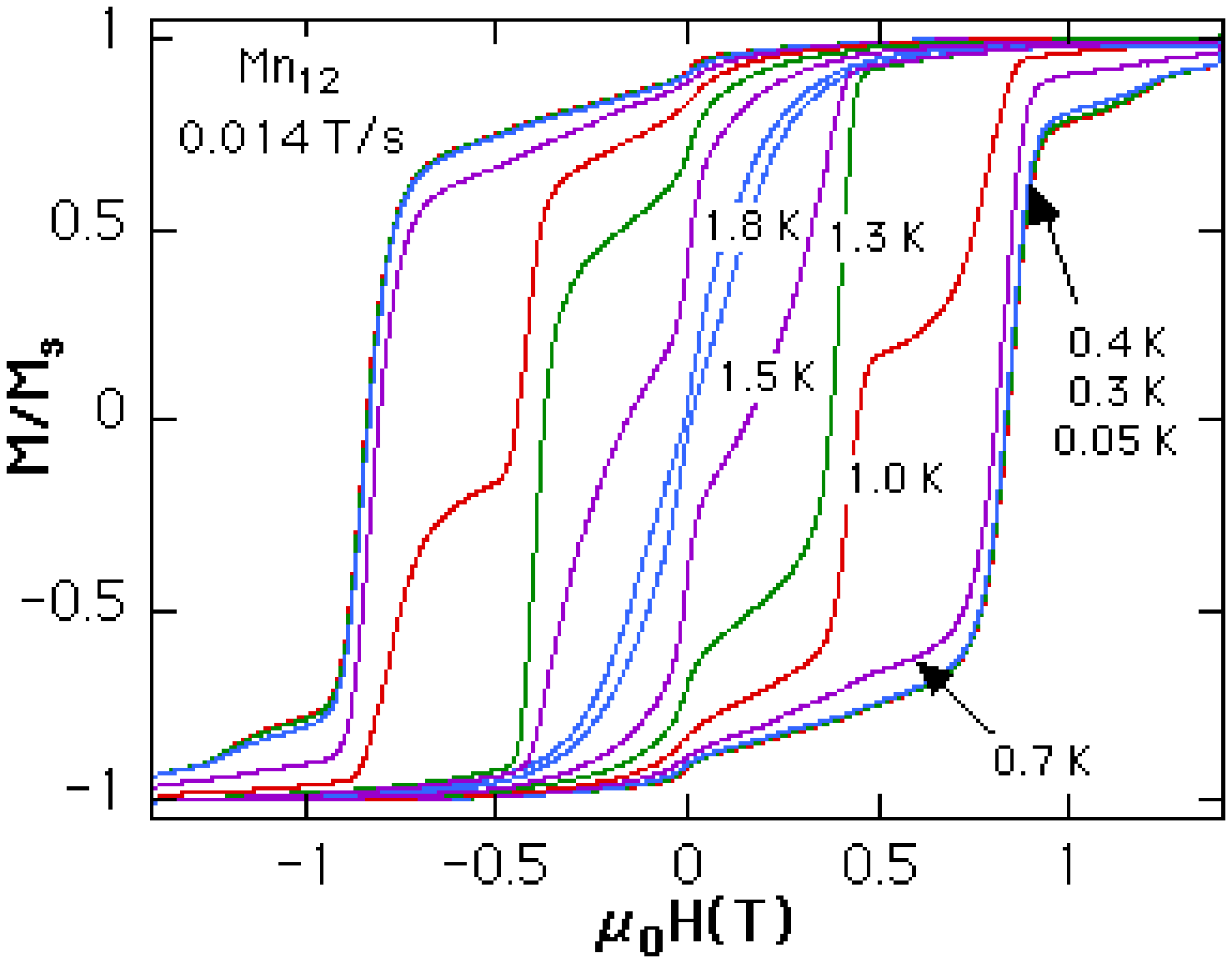}}
\centerline{\epsfxsize=8 cm \epsfbox{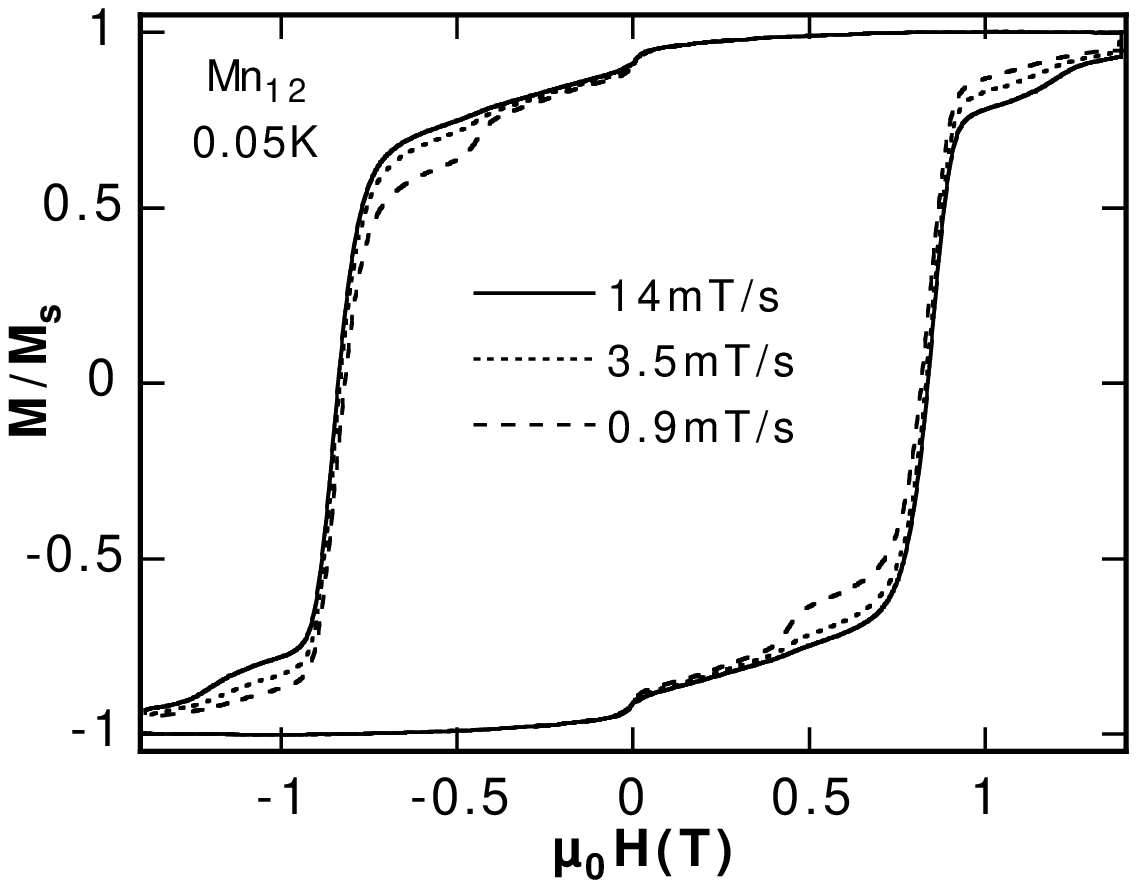}}
\caption{Hysteresis loops of a single crystal 
of [Mn$_{12}$]$^{2-}$ molecular clusters
at (a) different temperatures and a constant field
sweep rate and (b) 0.05 K and different field sweep
rates. Note the great difference of tunneling rate
between the resonance at 0.43 and 0.85 T that establish
the parity of the wave functions involved in the tunneling
process. This is quite different to Mn$_{12}$ acetate
showing a gradual increase of the tunneling propabilities.}
\label{hyst}
\end{figure}

Figs. 2a and 2b show typical hysteresis loop 
measurements on a single crystal of [Mn$_{12}$]$^{2-}$. 
The effect of avoided level crossings can 
be seen in hysteresis loop measurements. 
When the applied field is near an 
avoided level crossing, the magnetization relaxes faster, 
yielding steps separated by plateaus. 
As the temperature is lowered, there is a decrease 
in the transition rate as a result of 
reduced thermally assisted tunneling.
Below about 0.4 K, the  hysteresis loops become 
temperature independent which suggests that the
ground state tunneling is dominating. 
The field between two resonances
allows an estimation of the anisotropy constants $D$,
and a value of $D \approx $ 0.55 K was determined.

We have tried to use the
Landau--Zener method~\cite{Landau32,Zener32} to measure
the tunnel splitting as a function of transverse field
as previously reported for Fe$_8$~\cite{WW_Science99},. 
However, the tunnel probability in the pure quantum
regime (below 0.4 K) was too small for our measuring 
technique~\cite{note1}.
We therefore studied the tunnel probability in
the thermally activated regime~\cite{WW_EPL00}.

\begin{figure}
\begin{center}
\centerline{\epsfxsize=8 cm \epsfbox{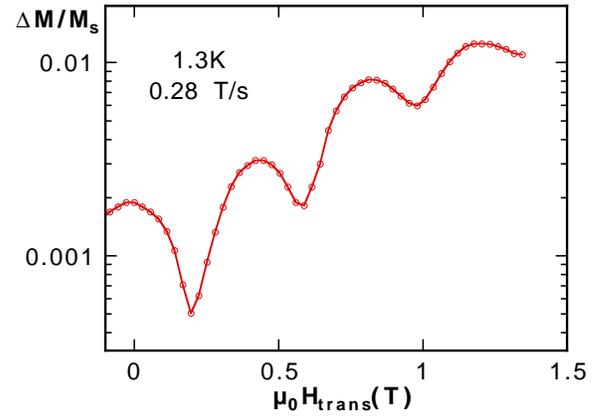}}
\caption{Fraction of molecules which reversed their magnetization after
the  field was swept over the zero field resonance.
The contribution of the  fast relaxing species
is substracted. The observed oscillations are 
direct evidence for quantum phase interference.}
\label{delta}
\end{center}
\end{figure}

In order to measure the tunnel probability,
a crystal of [Mn$_{12}$]$^{2-}$ SMM was first placed in a high 
negative field, yielding a saturated initial magnetization. 
Then, the applied field was swept at a constant rate 
over one of the resonance transitions and 
the fraction of molecules which 
reversed their spin was measured. This experiment was
then repeated but in the presence of a
constant transverse field. A typical result is
presented in Fig. 3 showing oscillations 
of the tunnel probability as a 
function of the magnetic field applied along the 
hard anisotropy axis. These oscillations are due to topological 
quantum interference of two tunnel paths of 
opposite windings~\cite{Garg93}. This observation is similar 
to the result on the Fe$_8$ molecular cluster~\cite{WW_Science99}. 
It is therefore the second direct evidence for 
the topological part of the quantum spin phase 
(Berry phase) in a magnetic system.
The period of oscillation
allows an estimation of the anisotropy constant $E$ (see Eq. 2) 
and a value of $E \approx $ 0.06 K was obtained.

In conclusion, magnetization measurements of a molecular clusters 
[Mn$_{12}$]$^{2-}$ with a spin ground state of $S = 10$ show
resonance tunneling at avoided energy level crossings. 
The observed oscillations of the tunnel probability as a 
function of a transverse field are due to topological 
quantum phase interference of two tunnel paths of 
opposite windings. [Mn$_{12}$]$^{2-}$ is therefore the
second molecular clusters presenting 
quantum phase interference.

%

\begin{thebibliography}{10}
\bibitem{note1} As observed for Mn$_{12}$ acetate~\cite{WW_EPL99},
a crystal of [Mn$_{12}$]$^{2-}$ contains a small fraction of
faster relaxing species, which are probably [Mn$_{12}$]$^{2-}$
having a defect. The signal of these species was very
large compared to the ground state relaxation rate
of the major species.
\end{thebibliography}

\newpage
\widetext
\end{document}